\documentstyle[aps,multicol,eqsecnum,epsf]{revtex}
\begin{document}
\draft
\title{\bf Scaling in Late Stage Spinodal Decomposition with Quenched
Disorder}

\author{Mark F.~Gyure,$^{1,2}$ Stephen T.~Harrington,$^1$ Richard
Strilka$^1$ and H.~Eugene Stanley$^1$}

\address{ $^1$Center for Polymer Studies, Department of Physics,
Boston University, Boston, MA 02215 \\ $^2$ Hughes Research
Laboratories, Malibu, CA 90265 }

\date{\today}
\maketitle

\begin{abstract}
We study the late stages of spinodal decomposition in a Ginzburg-Landau
mean field model with quenched disorder.  Random spatial dependence in
the coupling constants is introduced to model the quenched disorder.
The effect of the disorder on the scaling of the structure factor and on
the domain growth is investigated in both the zero temperature limit and
at finite temperature. In particular, we find that at zero temperature
the domain size, $R(t)$, scales with the amplitude, $A$, of the quenched
disorder as $R(t) = A^{-\beta} f(t/A^{-\gamma})$ with $\beta \simeq 1.0$
and $\gamma \simeq 3.0$ in two dimensions. We show that $\beta/\gamma =
\alpha$, where $\alpha$ is the Lifshitz-Slyosov exponent. At finite
temperature, this simple scaling is not observed and we suggest that the
scaling also depends on temperature and $A$.  We discuss these results
in the context of Monte Carlo and cell dynamical models for phase
separation in systems with quenched disorder, and propose that in a
Monte Carlo simulation the concentration of impurities, $c$, is related
to $A$ by $A \sim c^{1/d}$.
\end{abstract}

\vfill

\begin{multicols}{2}
\narrowtext

\section{Introduction}

Spinodal decomposition has received a great deal of attention in recent
years since it is relevant to many materials processing problems and is
one of the unsolved problems in statistical mechanics. Much progress has
occurred over the past decade in understanding both the early and late
time behavior of spinodal decomposition in the simplest systems with no
disorder and a single order parameter \cite{gunton,ising,gl,cell}, but
there is more to learn concerning the important process of spinodal
decomposition in disordered systems.  Systems with quenched disorder,
for example, are interesting because the kinetics of spinodal
decomposition differs markedly from the behavior of pure systems. In
particular, the usual power law growth of the characteristic domain
size, $R \sim t^{\alpha}$ is thought to change over to a logarithmic
growth ~\cite{huse}
\begin{equation}
R \sim (\ln t)^{\theta}.
\label{logt}
\end{equation}

The earliest computational studies of phase separation in systems with
quenched disorder were Monte Carlo [MC] simulations performed nearly a
decade ago \cite{grest85,srolovitz85,srolovitz87}.  MC simulations are
generally inefficient for studying the domain growth since diffusion of
particles through the system at late stages requires many updates.
However, MC simulations have the distinct advantage of introducing the
quenched disorder in a very natural way.  Using MC simulations, Grest
and Srolovitz \cite{grest85} found evidence of a scaling function
relating $R$ to the concentration of impurities in a two-dimensional
Ising model with nonconserved order parameter, but were unable to
confirm the prediction \cite{huse} of logarithmically slow domain
growth, Eq.~(\ref{logt}). Similar behavior was seen in MC simulations of
a model for simultaneous phase separation and gelation in an Ising-type
model with conserved order parameter \cite{ggscs}.

Further progress has recently been made \cite{puri91,puri92} using
more efficient cell dynamical methods \cite{oono}.  These cell
dynamical simulations were able to probe the late stage behavior for
both cases of conserved and nonconserved order parameter by
introducing the disorder as a phenomenological parameter in the
Ginzburg-Landau free energy ~\cite{MA}. Using cell dynamical methods,
Puri et al.  \cite{puri91,puri92} were able to observe the
logarithmically slow domain growth \cite{huse} for both conserved and
nonconserved order parameter, although the exponent $\theta$ in the
predicted behavior, Eq.  ~(\ref{logt}), for the case of conserved
order parameter, appeared to vary with the amplitude of the quenched
disorder.  This could be due in part to the systems not having reached
their asymptotic behavior; even using the cell dynamical approach,
simulation of disordered systems is computationally demanding due to
the logarithmically slow growth.

Here we study a time-dependent Ginzburg-Landau (TDGL) model and
consider scaling of the domain size as a function of the quenched
disorder both in the zero temperature limit and for $T>0$.  For
$T=0$, we expect that domain growth will effectively stop as various
parts of the system fall into metastable states from which they cannot
escape.  Although the $T=0$ limit is clearly unphysical, domain growth
pinning is seen in experimental systems with quenched disorder
\cite{bansil}.  Studying the TDGL model in this limit gives us the
opportunity to study the effect of a single parameter, the quenched
disorder $A$, on the dynamics of domain growth.  We find scaling that
relates the characteristic domain size $R$ to the phenomenological
parameter $A$ that describes the disorder in the model.  We also
examine the behavior of this model at finite temperature.  We find
that this simple scaling does not hold and propose a form for the
exponent $\theta$ in Eq.~(\ref{logt}).

The rest of this paper is organized as follows. In Section
\ref{sec:model}, we describe the time dependent Ginzburg-Landau model
with quenched disorder and briefly discuss its numerical
implementation. In Section \ref{sec:results}, we describe the results
of our simulations and propose a scaling form that is consistent with
these results. Finally in Section \ref{sec:discussion}, we place these
results in context with results for the Monte Carlo and cell dynamical
models described above.

\section{The Model and its Scaling Properties}
\label{sec:model}

In this section we briefly review the time dependent Ginzburg-Landau
(TDGL) equation and the rescaling of the variables that leads to the
Cahn-Hilliard equation \cite{cahn}, the basis of our simulation model.
In addition, we introduce the quenched disorder into the Cahn-Hilliard
equation \cite{puri91,puri92} via additional random, static coupling
constants taken from a uniform distribution of width $A$. We present a
scaling hypothesis for the model based on observations made in
previous work \cite{puri91,puri92}.

Consider a system with two independent species. The order parameter,
$\phi$, is defined \cite{gunton} to be the difference in the local
concentration of these two species

\begin{equation}
\phi \equiv c_{1} - c_{2}.
\label{orderp}
\end{equation}
For a conserved order parameter, the local concentration difference
obeys a continuity equation of the form
\begin{equation}
\frac{\partial}{\partial t} \phi({\bf r},t) = - {\bf \nabla} \cdot
{\bf j}({\bf r},t) + \zeta({\bf r},t).
\label{cont}
\end{equation}
Here the current is
\begin{equation}
{\bf j}({\bf r},t) = -M{\bf \nabla}\mu({\bf r}),
\label{current}
\end{equation}
where $M$ is the mobility (assumed to be constant), $\mu({\bf r})$ is the
chemical potential (the variational derivative of the free energy), and
$\zeta({\bf r},t)$ is Gaussian-distributed random noise that satisfies
\begin{equation}
\langle \zeta({\bf r},t),\zeta({\bf r}^{'},t^{'}) \rangle =
-2k_{B}TM~{\nabla}^2~\delta({\bf r}-{\bf r}^{'})~\delta(t-t^{'}),
\label{noise}
\end{equation}
where the brackets represent a thermal average over many different
configurations. This term guarantees that the proper equilibrium state
is achieved. In the TDGL model one usually chooses the Ginzburg-Landau
free energy functional \cite{gunton}
\begin{equation}
F = \int d{\bf r} \{ \frac{1}{2} \kappa {|\nabla \phi|}^2 - \frac{1}{2}
\rho {\phi}^2 +
\frac{1}{4}u{\phi}^4 \}
\label{free_energy}
\end{equation}
Combining the continuity equation, Eq.~(\ref{cont}), with the free
energy, Eq.~(\ref{free_energy}), yields
\begin{equation}
\frac{\partial \phi}{\partial t} = M{\nabla}^2 \{ -\kappa {\nabla}^2\phi -
\rho \phi + u{\phi}^3 \} + \zeta .
\label{modelB}
\end{equation}
The continuity equation, Eq.~(\ref{cont}), guarantees conservation of
the order parameter $\phi$; this is known as Model B.
Model B is a time-dependent Ginzburg-Landau model or relaxation model
since the time evolution of the order parameter depends only on the
minimization of the free energy functional, Eq.~(\ref{free_energy}).

The order parameter $\phi$ is redefined by scaling the phenomenological
parameters $M, \kappa, \rho$ and $u$ in Eq.~(\ref{modelB}) as well as
the dynamical variables ${\bf r}$ and $t$ to obtain a simplified
equation of motion for the rescaled order parameter. This is known as
the Cahn-Hilliard equation \cite{cahn}, and has the form
\begin{equation}
\frac{\partial}{\partial \tau} \psi = \frac{1}{2}{\nabla}^2(-{\psi} +
{\psi}^3 - {\nabla}^2\psi) + \sqrt{\epsilon}\xi .
\label{cahn-hilliard}
\end{equation}
The transformation that rescales Model B into the
Cahn-Hilliard model, following Grant et al. \cite{grant} and Rogers et
al. \cite{RED}, is
\begin{mathletters}
\begin{equation}
{\bf x} \equiv \left(\frac{\rho}{\kappa}\right)^{1/2}{\bf r},
\end{equation}
\begin{equation}
\tau \equiv \frac{2M{\rho}^2}{\kappa}t,
\end{equation}
\begin{equation}
\psi \equiv \left(\frac{u}{\rho}\right)^{1/2}\phi,
\end{equation}
\begin{equation}
\epsilon \equiv
{\frac{k_{B}Tu}{{\rho}^2}}\left({\frac{\rho}{\kappa}}\right)^{d/2}.
\end{equation}
\end{mathletters}
Hence from Eq.~(\ref{noise}),
\[\langle\xi({\bf x},\tau),\xi({\bf x}^{'},\tau^{'}) \rangle =\]
\begin{equation}
-2k_{B}Tu/{{\rho}^2}\left(\frac{\rho}{\kappa}\right)^{1/2}M~{\nabla}^2
{}~\delta({\bf x}-{\bf x}^{'})~\delta(\tau-\tau^{'}).
\end{equation}
\label{transform}
\noindent The formulation of model B in these variables clearly shows that
this model involves only a single parameter, $\epsilon$, proportional to
the temperature.

To incorporate quenched disorder into Model B, it is conventional to
introduce a spatial dependence in the parameters $\rho$ and $u$ of
Eq.~(\ref{free_energy}) ~\cite{MA}.  Consider couplings of the form
$\rho = \rho_{0} + \delta \rho({\bf x})$ and $u = u_{0} + \delta
u({\bf x})$, where $\rho_{0}$ and $u_{0}$ are identified with the free
energy in the absence of disorder.  After performing the
transformation of variables leading to Eq.~(\ref{cahn-hilliard}), the
time dependent Ginzburg-Landau equation with quenched disorder becomes
\[\frac{\partial}{\partial t} \psi =\]
\begin{equation}
-\frac{1}{2}{\nabla}^2\left[\left(1+\frac{\delta \rho({\bf
x})}{\rho_{0}}\right){\psi}-\left(1+\frac{\delta u({\bf
x})}{u_{0}}\right){\psi}^3 + {\nabla}^2\psi\right]+\sqrt{\epsilon}\xi.
\label{qdTDGL}
\end{equation}
We choose for simplicity the factors ~\cite{MA}
\begin{equation}
\frac{\delta\rho({\bf x})}{\rho_{0}}\psi,~~~~ \frac{\delta u({\bf
x})}{u_{0}}\psi^{3}
\label{disorder}
\end{equation}
to be random variables taken from a uniform distribution of width $A$, thereby
giving a random spatial dependence to the rescaled couplings in the model.

Equation~(\ref{qdTDGL}) is a generalization of the Cahn-Hilliard
equation, Eq.~(\ref{cahn-hilliard}).  Earlier results clearly indicate that
the crossover time, $t_{\times}$, from the early stage algebraic growth
to logarithmically slow growth at later times increases as the amplitude
$A$ of the quenched disorder decreases \cite{puri92}. Thus it is
appropriate to consider two distinct time domains separated by a
crossover region characterized by a time $t_{\times}$.  For $t \ll
t_{\times}$ we expect the domain size to grow as
\begin{equation}
R(t) \sim t^{\alpha},
\label{ls-scaling}
\end{equation}
where $\alpha = 1/3$ from the Lifshitz-Slyosov theory \cite{LS}. This
exponent has been observed in a wide variety of TDGL and cell dynamical
simulations \cite{gl,cell} with conserved order parameter. For $t \gg
t_{\times}$ we expect that the domain size at any time in the slower
growth region will scale as $R(t) \sim A^{-\beta}$.  Finally using the
observation by Puri and Parekh \cite{puri92} that the crossover time
$t_{\times}$ increases for decreasing disorder amplitude suggests that
\begin{equation}
t_{\times} \sim A^{-\gamma}
\label{scaling2}
\end{equation}
and hence the scaling form
\begin{equation}
R(t) \sim A^{-\beta} f(t/t_{\times}) \sim A^{-\beta} f(tA^{\gamma}),
\label{scaling1}
\end{equation}
{}From (\ref{ls-scaling}) and (\ref{scaling1}) it follows that
\begin{equation}
\beta = \alpha \gamma.
\label{exponents}
\end{equation}

\section{Calculations and Results}
\label{sec:results}

We now describe the results of our computer simulations that verify the
scaling hypothesis ~(\ref{scaling1}).  First we calculate the average
domain size, $R(t)$, of the system. Since there is only {\em one\/}
relevant length scale in a system undergoing spinodal decomposition
\cite{gunton}, any characteristic length, $R$, is proportional to the
domain size. We choose $R$ to be the inverse of the first moment $k_{1}$
of the spherically-averaged structure factor, $S(k,t)$, which is the
structure factor $S({\bf k},t) \equiv \langle
\int{d{\bf r} e^{-i{\bf k}\cdot({\bf r}-{\bf r'})}g({\bf r}-{\bf
r'})}\rangle$ averaged over the two-dimensional density of states. As
usual, $S({\bf k},t)$ is the Fourier transform of the density-density
correlation function $g({\bf r}-{\bf r'})$. The brackets represent the
averaging over many realizations of the disorder parameterized by $A$
(see Table~\ref{table1}).  Since we are considering systems with
conserved order parameter, the domain size scales as $R(t) \sim
1/k_{max}$, where $k_{max}$ is the value at which $S({\bf k},t)$ is
maximum.  In our simulations, however, we use the first moment $k_{1}$
of $S(k,t)$ (which is equivalent \cite{gunton}), since the peak maximum
is difficult to determine accurately.

We integrate the equation of motion Eq.~(\ref{qdTDGL}) directly, using
the central difference approximation for the spatial derivatives.
Following Rogers et al. \cite{RED}, we use a time step of $\Delta
t=0.15$ and lattice spacing $\Delta x=0.30$ to insure the stability of
Eq.~(\ref{qdTDGL}).  We allow for $3 \times 10^{6}$ updates for each run
on a two dimensional lattice of size $256 \times 256$ and calculate
S({\bf k},t) at logarithmically spaced time intervals.  We obtain the
results discussed in this section by averaging over $15$ to $25$
realizations of $A$ (see Table \ref{table1}), ranging from $0.0$ to
$0.45$.

We start with a homogeneous two-dimensional system with the rescaled
order parameter $\psi$ initialized from a uniform random distribution of
range $-0.005$ to $0.005$.  For $T>0$, we use a noise amplitude
$\sqrt{\epsilon}=0.5$, see Eq.~(\ref{qdTDGL}). In Fig.~\ref{fig1} we
display the value of $\psi$ for every choice of $A$.  Each snapshot is
made after $3\times 10^6$ updates in the time evolution.  As can be seen
from these snapshots, the larger the disorder parameter $A$, the smaller
the asymptotic domain size.  The domains gradually decrease in size as
we increase $A$.  This holds for simulations at both $T=0$, the top row,
and $T>0$, the bottom row.  Note that the domains have rougher and wider
interfaces for $T>0$.

Following the discussion in Sec.~\ref{sec:model}, we calculate the domain
size $R(t)$ from the first moment of the spherically-averaged structure
factor. In Fig.~\ref{fig2}a, we display our results for $S(k,t)$ vs.
wave number $k \equiv \mid{\bf k}\mid$, averaged over $20$ runs with no
quenched disorder. As expected,~\cite{gunton,RED},~$S(k,t)$ displays a
peak that increases with time to smaller values of wave number $k$.  The
peak of the structure factor is larger and occurs at a greater value of $k$
compared to the case with quenched disorder (see Fig.~\ref{fig2}b).  In
fact, the finite value of the asymptotic domain size for $A=0$ only
exists due to the finite size of the lattice.  This behavior is
contrasted with Fig.~\ref{fig2}b which shows $S(k,t)$ for $A=0.45$.  As
Fig.~\ref{fig2}b suggests, the peak of $S(k,t)$ reaches a maximum value
at a value of $k > 2\pi/L$.

Fig. \ref{fig3} shows the domain size $R(t)$ at (a) $T=0$ and (b) $T>0$
for each value of $A$.  As can be seen in Fig.~\ref{fig3}a,~$R(t)$
becomes constant after $10^5-10^6$ updates.  For $A=0.45$,~$R(t)$ is
constant after $2 \times 10^5$ updates, while the domain size for
$A=0.25$ is not constant even after $3\times10^6$ updates. However, we
expect that for even small values of $A$, $R(t)$ becomes constant.  This
behavior of $R(t)$ is called {\em pinning\/} and is a hallmark of
quenched disorder.  As we increase the amplitude of the quenched
disorder (increase the concentration of impurities), the domain growth
stops and eventually becomes pinned.  As Fig.~\ref{fig3}b shows, in the
case of $T>0$ the domains become larger for each value of $A$, but no
pinning behavior is observed.  In fact, the domains continue to grow
throughout the duration of the simulation, as expected from
\cite{huse}.  In both cases, two distinct time regimes are found: a
rapid growth in the domain size during the first $10^{3}-10^{4}$
updates, followed by a slower growth after some crossover region,
characterized by a time $t_{\times}$, that is dependent on the amplitude
of the quenched disorder.

Since the scaling hypothesis, Eq.~(\ref{scaling1}), was obtained from
considering the late stage growth behavior where $t \gg t_{\times}$, we
are interested in the scaling behavior of the asymptotic domain size.
Fig.~\ref{fig4} shows the asymptotic values $R_{f}$ for each value of
$A$ for (a) $T=0$ and (b) $T>0$.  We find a linear relation, confirming
the first part of the scaling hypothesis, namely that the domain size at
late times scales as $R \sim A^{-\beta}$ with $\beta \simeq 1$. We also
consider scaling of the crossover time, $t_{\times}$, for each $A$.  We
assume that the crossover time scales as $t_{\times} \sim A^{-\gamma}$,
where $\gamma \simeq 3$ from Eq.~(\ref{exponents}). Fig.~\ref{fig5}
shows the simulation results for $R(t)$ rescaled by Eq.~(\ref{scaling1})
with $\beta=1, \gamma=3$.

Notice that the zero temperature result, shown in Fig.~\ref{fig5}a,
collapses quite well over four decades in rescaled time, clearly
confirming the scaling hypothesis. The finite temperature result,
Fig.~\ref{fig5}b, scales well over the early time region only, with
a clear breakdown in scaling at late times. This suggests that the
exponent $\theta$ in Eq.~(\ref{logt}) has a functional dependence on
the amplitude of the quenched disorder and the temperature
\begin{equation}
\theta=\theta(A,T).
\label{theta}
\end{equation}
We discuss reasons for this in the following section.

\section{Discussion}
\label{sec:discussion}

We find that a time-dependent Ginzburg-Landau model with quenched
disorder is capable of reproducing similar results to those found in
earlier Monte Carlo simulations
\cite{grest85,srolovitz85,srolovitz87}, more recent cell dynamical
simulations \cite{puri91,puri92,oono}, and other TDGL simulations
\cite{sbs}. Our main interest here is the determination of the scaling
function relating the domain size $R$ to the amplitude of the disorder
$A$. For $T=0$, we find scaling similar to that proposed by Grest and
Srolovitz \cite{grest85} as well as Glotzer et. al.  \cite{ggscs} in MC
simulations.  Grest and Srolovitz find $\alpha = 1/2$ while Glotzer et.
al. find $\alpha = 1/4$ rather than the $1/3$ observed in our TDGL
simulations.  The reason for the different values of $\alpha$ is that MC
simulations rarely see the predicted Lifshitz-Slyozov growth exponent of
$1/3$ since particle diffusion in an MC simulation takes many more
updates than in a TDGL simulation. We expect that at low temperatures a
MC simulation using an accelerated algorithm \cite{fastMC} or simply
allowed to run sufficiently long will find a similar scaling function
with $\alpha = 1/3$.

Nevertheless, comparison of the scaling functions for the zero
temperature models described above enables us to find a relationship
between the phenomenological disorder parameter $A$ and the
concentration $c$ of quenched impurities used in Monte Carlo
simulations. The scaling form suggested by MC simulations is $R =
c^{-\beta'} f(t c^{\gamma'})$ with $\beta' = 1/d$ \cite{ggscs}. The
scaling form suggested by our simulations is given by
Eq.~(\ref{scaling1}) as $R = A^{-\beta} f(t A^{\gamma})$ with $\beta
\simeq 1$. Since we expect the domain size to scale the same way in
both models, we anticipate the relationship
\begin{equation}
A \sim c^{1/d}
\label{scaling}
\end{equation}
between the amplitude of the quenched disorder and the impurity
concentration. This result indicates that direct MC simulations can be
replaced by TDGL simulations or even more efficient cell dynamical
simulations, since the relationship between the amplitude of the
quenched disorder can now be understood directly in terms of the
impurity concentration for a given model.

We note that the scaling found in the simulations discussed above was
observed either at zero temperature \cite{grest85} or effectively at
zero temperature \cite{ggscs}. As our results in Fig.~(\ref{fig5}b)
show, the scaling behavior for quenched disorder models at finite
temperature is more complicated. Puri and Parekh \cite{puri92} observe
that the asymptotic behavior of their cell dynamical model was not
independent of the disorder parameter $A$. Similarly, the MC results of
Grest and Srolovitz at $T>0$ also suggests that the asymptotic behavior
depends on the concentration of impurities. Thus we do not expect that
the simple scaling hypothesis, Eq.~(\ref{scaling1}), would be valid for
$T>0$. For late stage domain growth, we expect that the exponent
$\theta$, in Eq.~(\ref{logt}), is a function of $A$ and $T$.  However, a
scaling function describing both the finite and zero temperature
behavior of quenched disorder models might still exist.  A detailed
study of one or more of these models (MC, TDGL, or cell dynamical)
spanning a range of temperature and disorder parameters would likely
shed considerable light on this matter. Such a study would involve a
large computational effort, and is well beyond the scope of the present
work.

Finally, we suggest that the type of scaling behavior found in a
variety of quenched disorder models should be experimentally
observable. Indeed, experiments could test the validity of the TDGL
model discussed here, and could provide interesting insights into the
universality of the scaling we report.

We thank L.~A.~N.~Amaral, A.~Coniglio, R.~Cuerno, S.~C.~Glotzer,
S.~Havlin, S.~Sastry and G.~Valley for helpful discussions.  The Center
for Polymer Studies is supported in part by funds from the NSF, and STH
is supported by an NIH doctoral training fellowship.  All simulations
were done using the 32-node CM-5 at Boston University's Center for
Computational Science.

\begin{table}
\caption{Asymptotic values, $R_{f}$, of the domain size for various values
of the quenched disorder as calculated from the simulations for
both $T=0$ and $T>0$.  We used a $256$x$256$ lattice
with spacing $\Delta x=0.30$.  The evolution of Eq.~(2.9) was performed
for $3$x$10^6$ updates with a time interval of $0.15$.}
\vspace{2cm}
\begin{tabular}{c|cc|cc}
     &       ~~~~~$T=0$  & & ~~~~~$T>0$ & \\
Amplitude     & Runs & $R_{f}$  & Runs & $R_{f}$  \\
\tableline
    $0.25$ & $23$ & $6.2\pm 0.1$  &  &   \\
    $0.30$ & $26$ & $5.26\pm 0.07$  &  $7$ & $7.4\pm 0.1$   \\
    $0.35$ & $15$ & $4.45\pm 0.05$  & $10$ & $6.4\pm 0.1$   \\
    $0.40$ & $18$ & $3.92\pm 0.04$  &  $8$ & $5.3\pm 0.2$    \\
    $0.45$ & $18$ & $3.50\pm 0.03$  &  $6$ & $4.6\pm 0.1$   \\
\end{tabular}
\label{table1}
\end{table}

\begin{figure}
  \caption{Typical simulation results showing the domains for various
values of $A$ after $3 \times 10^6$ updates.
The top row is the case $T=0$ and the bottom row is $T>0$.  Note the
decreasing domain size as $A$ increases.  Also, the interface between
the two phases is thicker and rougher for $T>0$ compared to $T=0$.}
\label{fig1}
\end{figure}

\begin{figure}
  \caption{(a) Spherically-averaged structure factor $S(k,t)$ vs.
wave number $k$ for $A=0$
at various times. The peak value increases with time and approaches
$k=0$. (b) $S(k,t)$ vs. $k$ for $A=0.45$. Note that $S(k,t)$ is
smaller, and $k_{max}$ is larger, than in (a). }
\label{fig2}
\end{figure}

\begin{figure}
  \caption{(a) Double logarithmic plot of the average domain size $R(t)$
vs. time $t$ at $T=0$ for several values of $A$.  The pinning regime is
clearly reached for all disorder amplitudes except, possibly, for
$A=0.25$.  The $t^{1/3}$ behavior is apparent at short times, while $R
\sim A^{-1}$ is found in the late stages. (b) $R(t)$ vs. $t$ for $T>0$
for several values of $A$.  Note that pinning is not observed for any
$A$ .} \label{fig3}
\end{figure}

\begin{figure}
  \caption{(a) Inverse of the asymptotic domain size $1/R_{f}$ vs.
disorder amplitude $A$ for $T=0$.  $R_{f}$ is taken to be the average
domain size after $3\times10^6$ updates.  The linear relation confirms
the scaling hypothesis Eq.~(2.13) with $\beta=1$.  (b) $1/R_{f}$ vs. $A$
for $T>0$.  Note the deviations from a linear relation.}
\label{fig4}
\end{figure}

\begin{figure}
  \caption{(a) Scaled domain size $RA$ vs. scaled time $tA^{3}$ for
$T=0$, using $\beta=1$ and $\gamma=3$ in the scaling relation
Eq.~(2.13).  The collapse onto a single curve further confirms our
scaling hypothesis. (b) Same plot as (a) using simulations with $T>0$.}
\label{fig5}
\end{figure}

\end{multicols}

\end{document}